\documentclass[prb,preprint]{revtex4-1}

\usepackage{graphicx}
\usepackage{amsmath}
\usepackage{amssymb}
\usepackage{dcolumn}

\newcommand{\dT}{\Delta T}
\newcommand{\dL}{\Delta L}
\newcommand{\dLt}{\dL_\text{t}}
\newcommand{\dLm}{\dL_\text{e}}

\newcommand{\units}{$\mu$m/m$\cdot$K}

\newcommand{\FIGapparatus}{
\begin{figure}[t]\center
\includegraphics[width=\columnwidth]{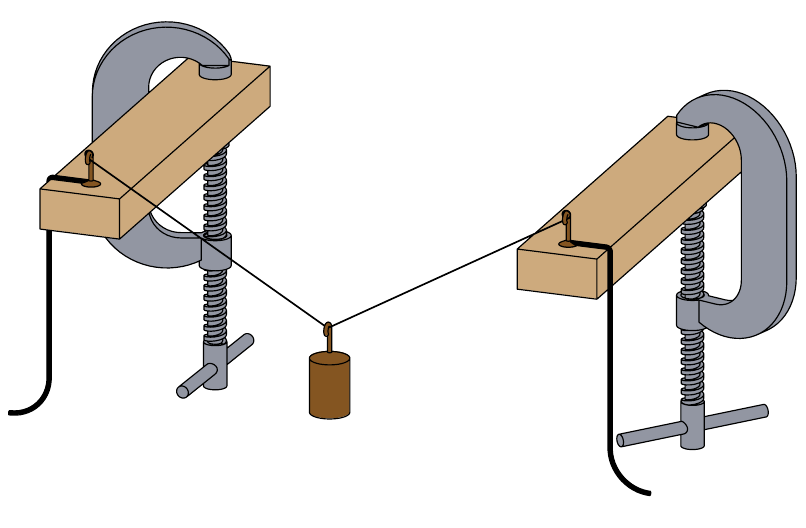}
\caption{\label{fig:apparatus} Experimental apparatus. The ends of a wire are attached to blocks of wood via small metal hooks. Both blocks are fastened to a level, rigid surface (not shown) via c-clamps. Insulated copper wires are attached to the hooks to facilitate electrical connections to an ac power supply, as shown in Fig.~\ref{fig:schematic}. Our apparatus is an adaptation of similar setups used in Refs.~[9] and [10].}
\end{figure}
}

\newcommand{\FIGschematic}{
\begin{figure}[t]\center
\includegraphics[width=\columnwidth]{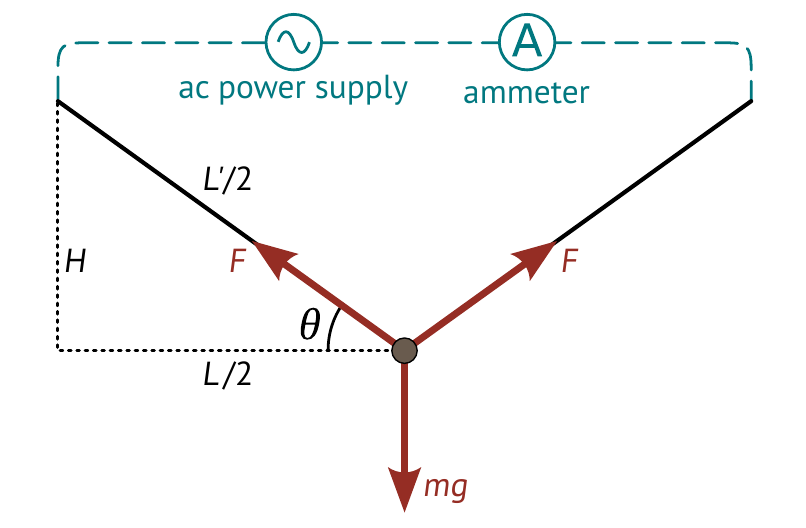}
\caption{\label{fig:schematic} Schematic of experimental setup. A load (filled circle) is suspended from the midpoint of a taut wire (solid black lines). The wire is connected in series with an ac power supply and an ammeter. Dashed blue lines indicate electrical connections and thick red arrows represent the forces experienced by the load.}
\end{figure}
}

\newcommand{\FIGcatenary}{
\begin{figure}[t]\center
\includegraphics[width=\columnwidth]{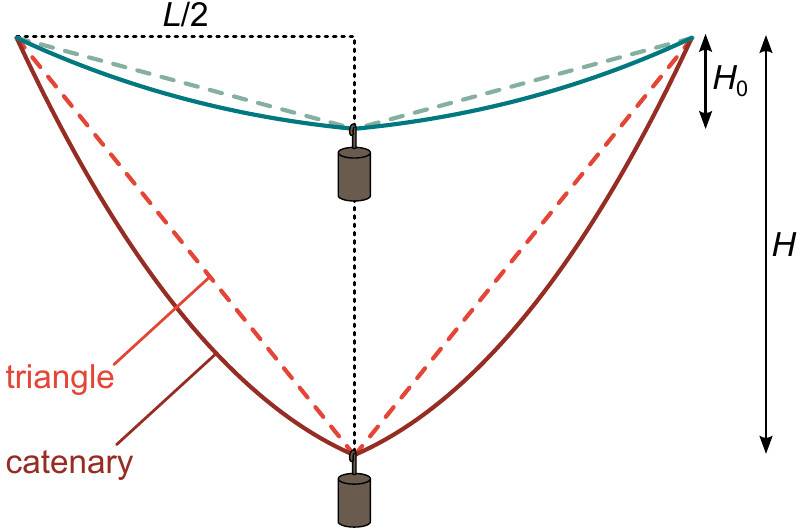}
\caption{\label{fig:catenary} Effects of the wire's elasticity and mass.  When the wire is at room temperature (upper, blue lines), the displacement $H_0$ of the load is nonzero due to elastic stretching of the wire. The displacement $H$ of a hot wire (lower, red lines) is due to a combination of elastic and thermal effects. In the massless wire approximation, the wire hangs in a triangle (dashed lines). A massive wire, on the other hand, hangs in the shape of a loaded catenary (solid lines).}
\end{figure}
}

\newcommand{\FIGspringdata}{
\begin{figure}[t]\center
\includegraphics[width=\columnwidth]{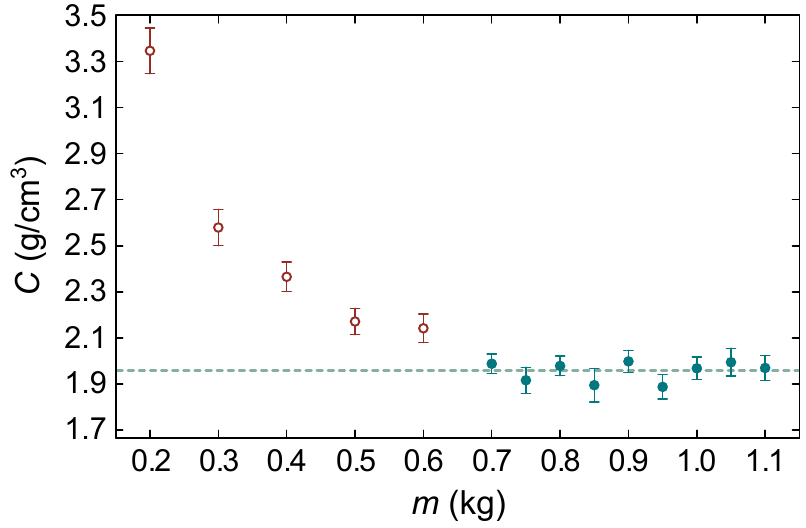}
\caption{\label{fig:k} Empirical determination of the massless wire regime. Shown are measurements of the quantity $C\equiv m/H_0^3$ as a function of $m$, where $H_0$ and $m$ are the room-temperature displacement and mass of the load, respectively. We use the dependence of $C$ on $m$ as an indicator of the validity of the massless wire assumption. When $C$ is constant with respect to $m$ the wire's mass is negligible; otherwise it is not. For our apparatus, the massless wire regime is realized when $m\gtrsim 0.7$~kg. The dashed line represents the weighted average of the data collected in this regime (filled blue circles). Error bars in this and subsequent figures represent statistical uncertainties.}
\end{figure}
}

\newcommand{\FIGthermaldata}{
\begin{figure}[t]\center
\includegraphics[width=\columnwidth]{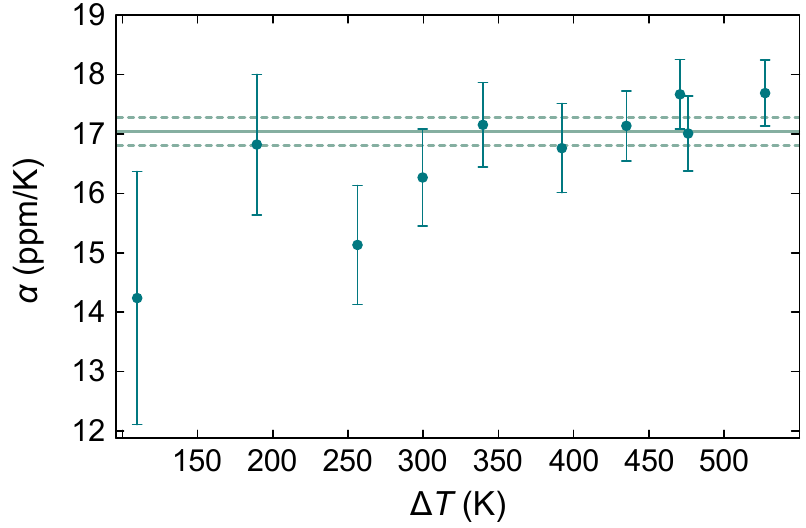}
\caption{\label{fig:alpha} Experimental determination of nichrome's thermal expansion coefficient $\alpha$ as a function of the temperature difference $\dT$ of the wire relative to room temperature. The data were analyzed using the massless, elastic wire model of Section~\ref{sec:k}. The solid line represents the weighted mean of the data, and the dashed lines represent the 65\% confidence interval due to statistical uncertainties. When taking systematic uncertainties into account (Table~\ref{tab:budget}), we find  $\alpha=17.1(1.3)$~\units, which is consistent with the accepted value of nichrome's thermal expansion coefficient, 17.3~\units.~\cite{Shackelford2001}}
\end{figure}
}

\newcommand{\FIGX}{
\begin{figure}[t]\center
\includegraphics[width=\columnwidth]{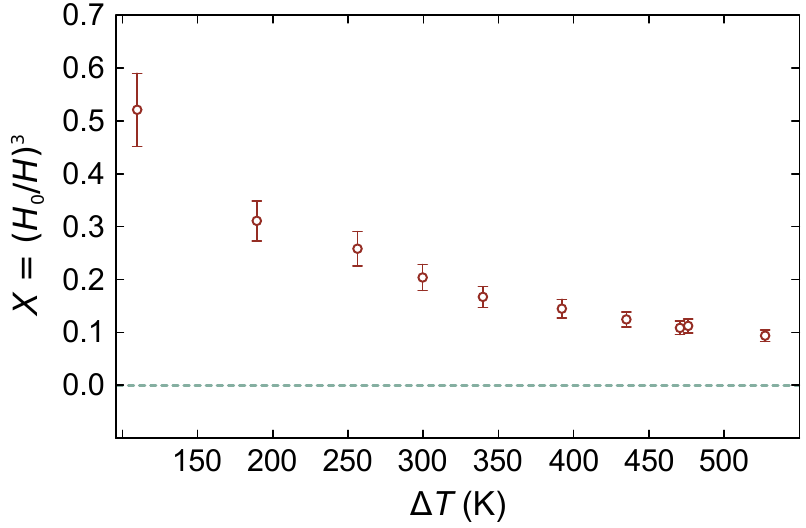}
\caption{\label{fig:X} Empirical determination of the stiff wire regime. Shown are measurements of the dimensionless quantity $X\equiv (H_0/H)^3$ as a function of temperature change $\dT$. When $X$ is negligibly small, the wire's elasticity can be neglected. For comparison, a dashed line corresponding to $X=0$ has been included. Because our data yield nonzero measurements of $X$ over a broad range of temperatures, the stiff wire regime is not realized in our experiment.}
\end{figure}
}

\newcommand{\TABbudget}{
\begin{table}
\caption{\label{tab:budget}Partial uncertainties in determination of $\alpha$.}
\begin{ruledtabular}
\begin{tabular}{ld}
Source of uncertainty & \multicolumn{1}{c}{Uncertainty in $\alpha$ (\%)} \\ \hline
Systematic: \\
\hspace{10pt}Displacement of load & 7.1 \\
\hspace{10pt}Length of wire & 2.3 \\
Statistical: \\
\hspace{10pt}Displacement of load & 1.4 \\
\hspace{10pt}Temperature change & 0.2 \\ \hline
\textbf{Total (added in quadrature)} & \textbf{7}.\textbf{6}
\end{tabular}
\end{ruledtabular}
\end{table}
}

\begin{document}

\title{Uncertainty Analysis for a Simple Thermal Expansion Experiment}
\author{Dimitri~R.~Dounas-Frazer}
\email{drdf@berkeley.edu}
\affiliation{Department of Physics, University of California at
Berkeley, Berkeley, California 94720, USA}
\author{Geoffrey Z. Iwata}
\email{gzi2000@columbia.edu}
\affiliation{Department of Physics, University of California at
Berkeley, Berkeley, California 94720, USA}
\affiliation{Department of Physics, Columbia University, New York, New York 10027, USA}
\author{Punit R. Gandhi}
\email{punit\_gandhi@berkeley.edu}
\affiliation{Department of Physics, University of California at
Berkeley, Berkeley, California 94720, USA}
\date{\today}

\begin{abstract}
We describe a simple experiment for measuring the thermal expansion coefficient of a metal wire and discuss how the experiment can be used as a tool for exploring the interplay of measurement uncertainty and scientific models. In particular, we probe the regimes of applicability of three models of the wire: stiff and massless, elastic and massless, and elastic and massive. Using both analytical and empirical techniques, we present the conditions under which the wire's mass and elasticity can be neglected. By accounting for these effects, we measure nichrome's thermal expansion coefficient to be $17.1(1.3)$~\units, which is consistent with the accepted value at the 8\% level.
\end{abstract}

\maketitle

\section{Introduction}

What does it mean for an effect to be negligible? This was the overarching question of a course we designed and taught to freshmen in their second semester at UC Berkeley through the Compass Project.~\cite{Albanna2012} To answer it, students must develop a sophisticated understanding of two important, interrelated physics concepts: measurement uncertainty and models. We used a thermal expansion experiment as a tool for facilitating this understanding. The phenomena relevant to the experiment (thermal expansion, elastic stretching, tension, and gravity) are familiar to students with only an introductory physics background. Thus this simple, low-cost experiment provides an accessible context for beginning students to tackle an abstract and sophisticated question about physics, \emph{i.e.}, what it means for an effect to be negligible. In this work, we present a detailed description of the experiment.

Thermal expansion is ubiquitous in our everyday lives, playing an important role in everything from rising sea levels~\cite{Rahmstorf2007} to the design of bridges~\cite{Moorty1992} and the performance of steel beam structures during fires.~\cite{Yin2005a,*Yin2005b}  Unsurprisingly, experiments and demonstrations for teaching about thermal expansion abound. Recent examples  include optical measurements of the expansion of copper,~\cite{Graf2012, Scholl2009} qualitative demonstrations of the contraction of rubber,~\cite{Liff2010} and techniques for measuring thermal expansion of very cold two-dimensional samples.~\cite{Carles2005} We focus on a previously proposed thermal expansion experiment detailed in Refs.~[9, 10]. A wire is pulled taut between two fixed anchors and a hanging mass is attached to its midpoint (Fig.~\ref{fig:apparatus}). As the wire is heated, it expands,~\cite{Hitchcock1945, Insley1984} causing the hanging mass to drop lower to the ground. By measuring the change in height of the mass, one can determine the change in length of the wire and hence its coefficient of thermal expansion.

We analyze two important sources of uncertainty that affect this experiment but have been ignored by previous studies:~\cite{Trumper1997, Liem1987} the mass and elasticity of the wire. To account for their contributions to the height of the load, we explore different models of the wire, including an idealized model which assumes a stiff, massless wire and a more realistic one which treats the wire as both massive and elastic. We describe how the wire's mass and elasticity affect measurement of the thermal expansion coefficient and determine the conditions under which their effects are negligible.

\section{Models of the wire}

We aim to determine the thermal expansion coefficient $\alpha$ of a wire with length $L$ at room temperature $T$. Ignoring effects that are nonlinear in temperature, the expansion coefficient satisfies~\cite{Giancoli1991}
\begin{equation}\label{eq:thermal}
\dLt = \alpha \dT L,
\end{equation}
where $\dLt$ is the change in length of the wire due to an increase in its temperature by $\dT$. For lengths and temperatures that are easy to achieve in a classroom, $\dLt$ is very small. Therefore, we use the geometry in Fig.~\ref{fig:schematic} to amplify the effects of thermal expansion.~\cite{Trumper1997, Liem1987} The wire is stretched horizonatally between two anchors and pulled taut by hand. A load of mass $m$ is attached to its midpoint. The wire expands as it heated, causing the load to drop lower to the ground by a distance $H$. A small change in the wire's length corresponds to a relatively large $H$. For example, increasing the temperature of a 175 cm wire by 100~K results in only a 3~mm change in length but causes the load to drop by about 7~cm. The tradeoff for this
order-of-magnitude increase in the size of the effect is that the mass and elasticity of the wire lead to systematic errors in determination of $\alpha$. Studying these two sources of uncertainty is one of the major goals of the present work.

\FIGapparatus

In this section, we present three models of the wire: stiff and massless, elastic and massless, and elastic and massive. We also outline analytical and empirical methods for determining when the effects of the wire's elasticity and mass are negligible.

\subsection{Stiff, massless wire}\label{sec:stiff}

The simplest model of the wire is one that neglects its mass and elasticity.  This model is valid when the wire is sufficiently stiff that elastic stretching is negligible compared to thermal expansion, and when the mass of the load is much larger than that of the wire. As we discuss in Section~\ref{sec:wiremass}, modeling the wire as massless is equivalent to assuming that it hangs in the shape of a triangle~(Fig.~\ref{fig:schematic}). Therefore, the change in length $\Delta L$ of the wire is related to the vertical displacement $H$ of the load by the Pythagorean Theorem:
\begin{equation}\label{eq:geomean}
H=\sqrt{L\dL/2},
\end{equation}
where terms of order $\dL^2$ have been neglected since $\left(\dL/L\right)^2\simeq 10^{-6}$ is negligible compared to the precision of $10^{-2}$ achieved in this experiment (Section~\ref{sec:results}). As Eq.~(\ref{eq:geomean}) shows, $H$ is proportional to the geometric mean of $L$ and $\dL$. Hence the displacement of the load is much larger than the actual change in length of the wire.

By modeling the wire as stiff, we assume that its elongation is due only to thermal expansion, \emph{i.e.}, $\dL=\dLt$. Solving Eqs.~(\ref{eq:thermal}) and (\ref{eq:geomean}) for the coefficient of thermal expansion yields~\cite{Trumper1997, *Trumper1998}
\begin{equation}\label{eq:alphaSM}
\alpha_0=\frac{1}{\dT}\frac{\dL}{L}=\frac{2}{\dT}\left(\frac{H}{L}\right)^2,
\end{equation}
where the subscript ``0" is included to distinguish Eq.~(\ref{eq:alphaSM}) from the results of the elastic wire model.

\FIGschematic

\subsection{Elastic, massless wire}\label{sec:k}

The effects of elastic stretching are apparent at room temperature~(Fig.~\ref{fig:catenary}): the wire stretches under the weight of the load, resulting in a nonzero displacement $H_0$ when $\dT=0$.  Therefore, Eqns.~(\ref{eq:thermal}) and (\ref{eq:geomean}) imply that $\dL\neq\dLt$.   We present a more realistic model that accounts for elastic stretching of the wire. In this case, there are two contributions to the wire's change in length:
\begin{equation}\label{eq:dL}
\dL = \dLt + \dLm,
\end{equation}
where $\dLt$ and $\dLm$ are the contributions from thermal and elastic effects, respectively. To model elastic stretching, we rely on Hooke's Law:~\cite{Giancoli1991}
\begin{equation}\label{eq:Hooke}
F=k\dLm,
\end{equation}
where $F$ is the tension in the wire and $k$ is the wire's spring constant, which we assume to be independent of temperature.

Thermal expansion and elastic stretching can be discriminated from one another based on their different scaling with $H$. We solve for $\dLt$ and $\dLm$ in terms of $H$ as follows. For a massless wire, the vertical component of the tension in the wire must exactly balance the weight of the load:
\begin{equation}\label{eq:balance}
2F\sin\theta=mg,
\end{equation}
where $\theta$ is defined in Fig.~\ref{fig:schematic}. Combining Eqs.~(\ref{eq:Hooke}) and (\ref{eq:balance}) and using the small angle approximation $\sin\theta\approx\tan\theta=2H/L$, we find
\begin{equation}\label{eq:dLm}
\dLm=mgL/(4kH).
\end{equation}
Substituting Eqs.~(\ref{eq:dL}) and (\ref{eq:dLm}) into (\ref{eq:geomean}) and solving for $\dLt$ yields
\begin{equation}\label{eq:dLt}
\dLt = 2H^2/L-mgL/(4kH).
\end{equation}
Finally, solving Eqs.~(\ref{eq:dLt}) and (\ref{eq:thermal}) for $\alpha$ gives
\begin{equation}
\label{eq:alphaEM}
\alpha = \frac{2}{\dT}\left(\frac{H}{L}\right)^2\left[1-\left(\frac{H_0}{H}\right)^3\right].
\end{equation}
The displacement $H_0$ is due only to elastic stretching of the wire and is related to the spring constant by
\begin{equation}\label{eq:k}
k = \frac{1}{8}\frac{mgL^2}{H_0^3}.
\end{equation}
Equation~(\ref{eq:k}) follows from Eqs.~(\ref{eq:geomean}) and (\ref{eq:dLm}) when $\dT=0$ and hence $\dL=\dLm$ and $H=H_0$. Thus, by measuring $H_0$, Eqs.~(\ref{eq:alphaEM}) and (\ref{eq:k}) allow us to account for elastic stretching when determining the thermal expansion coefficient and to measure the spring constant of the wire in a straightforward way.

To understand the conditions under which the elastic wire model presented here reduces to the stiff model presented in Section~\ref{sec:stiff}, we define the following dimensionless parameter:
\begin{equation}\label{eq:X}
X\equiv \left(\frac{H_0}{H}\right)^3 \approx \frac{mg}{kL}\left(\frac{1}{2\alpha\dT}\right)^{3/2}.
\end{equation}
The approximation, which is valid when $\dLm\ll\dLt$ and $\alpha\approx\alpha_0$, follows from solving Eqs.~(\ref{eq:alphaSM}) and (\ref{eq:k}) for $H$ and $H_0$, respectively. Writing Eq.~(\ref{eq:alphaEM}) as $\alpha=\alpha_0(1-X)$ shows that $X$ represents the fractional correction to $\alpha_0$ due to the wire's elasticity. When $X$ is negligibly small, so, too, are the effects of elastic stretching. For a particular wire with fixed $L$, $k$, and $\alpha$, such a regime is realized for sufficiently large temperature differences; as $\dT$ increases, elongation due to thermal expansion relaxes the wire tension and hence decreases $\dLm$. Although decreasing the load's mass has a similar effect, our analysis may not be valid if $m$ is too small because the massless wire model is only applicable when the load is very heavy compared to the wire.

\FIGcatenary

\subsection{Elastic, massive wire}\label{sec:wiremass}

To take the wire's mass into account, we model the shape of the hanging wire as an elastic, loaded catenary (Fig.~\ref{fig:catenary}). In the limit that the load is much heavier than the wire, the loaded catenary closely resembles a triangle and Eq.~(\ref{eq:geomean}) is valid. However, in general, numerical analysis is needed to determine how the length of the loaded catenary depends on the displacement and mass of the load, and the mass and elasticity of the wire.~\cite{Zapolsky1990, Irvine1976}

Rather than use a numerical approach to probe the regime of applicability of the elastic, massless wire model, we adopt an empirical one. One of the predictions of that model is that the quantity
\begin{equation}
C\equiv m/H_0^3
\end{equation}
is independent of $m$; indeed, Eq.~(\ref{eq:k}) implies that $C$ is constant and equal to $8k/(gL^2)$. However, there are two limits in which $C$ is not constant: when the load is too light, the wire cannot be modeled as massless and Eq.~(\ref{eq:k}) is not valid, and; in the opposite limit of a very heavy load, the wire may undergo nonlinear elastic deformation or it may be permanently stretched through plastic deformation, invalidating Hooke's Law~(\ref{eq:Hooke}). The quantity $C$ deviates from its constant value in both cases. We therefore use the scaling of $C$ with $m$ as an indicator for the breakdown of the elastic, massless wire approximation: for light loads, an observed dependence of $C$ on $m$ is evidence that the wire's mass cannot be neglected; for heavy loads, such a dependence indicates a breakdown of Hooke's Law.

\section{Results and Discussion}\label{sec:results}

\FIGspringdata

Our apparatus (Fig.~\ref{fig:apparatus}) consisted of: a 24~AWG nichrome wire with mass of 3~g and length of 175~cm; a set of hanging weights with masses 0.05, 0.1, 0.2, 0.5, and 1.0~kg; a 400~W variable ac power supply capable of supplying up to 9~A (rms) of current; a digital handheld thermocouple probe thermometer (Omega Engineering model HH11B); two wooden blocks (2"$\times$1"$\times$12") with metal hooks; two c-clamps; electrical wires; an ammeter; a ruler with millimeter precision; and a small mirror.

The ends of the wire were attached to the hooks on the wooden blocks. The blocks were secured to a table with the c-clamps, pulling the wire taut in the process. A load of variable mass was achieved by attaching different combinations of the hanging weights to the wire's midpoint. We affixed the ruler to the edge of the table about 1.5~cm behind the center of the wire and placed the mirror behind the ruler to minimize parallax error. Each end of the wire was connected to the ac power supply~(Fig.~\ref{fig:schematic}), and current was run through the wire causing its temperature to increase due to resistive heating. By varying the current from 0 to 9~A, we achieved temperature differences of up to 550~K. The large currents and high temperatures involved in the experiment warrant various safety precautions, such as installing a 10~A fuse on the power supply and exercising care near the hot wire. We measured the temperature of the wire in 4 places spaced 35~cm apart from each other and from the ends of the wire.

\FIGthermaldata

We performed two experiments. The first experiment was an empirical probe of the limits of the massless wire model by measuring the displacement $H_0$ of the load at room temperature as a function of load mass $m$ (Section~\ref{sec:wiremass}). The second experiment was a determination of nichrome's thermal expansion coefficient $\alpha$ by measuring the displacement $H$ of the load as a function of the wire's temperature change $\dT$. For this second measurement, data were analyzed according to Eqs.~(\ref{eq:alphaEM}) and (\ref{eq:k}), which are valid when the wire is massless and elastic (Section~\ref{sec:k}). For a given $m$ and $\dT$, measurements of $H$ and $H_0$ were repeated 3 to 5 times and we assumed the statistical uncertainty in determination of the displacement was given by the standard error of the mean of the repeated measurements. Statistical uncertainties were scaled to give a reduced chi-squared of unity when computing averages; scale factors were on the order of unity, indicating a good fit between models and data. Uncertainties in the calculated quantities (\emph{e.g.}, $\alpha$) were determined using standard error propagation methods.~\cite{Taylor1997}

We measured the displacement of the load relative to the position of the midpoint of the taut, unloaded wire. This choice of reference for the displacement leads to a systematic uncertainty of about $\pm5$~mm in $H$ and $H_0$ due to sag and kinks in the unloaded wire, which is the dominant source of uncertainty in determination of $k$ and $\alpha$ (Table~\ref{tab:budget}). Future experiments should be improved by using a more accurate reference from which to measure the load's displacement.

\TABbudget

Systematic effects that lead to gradual increase of the load's displacement over time may introduce biases in our measurements. One possible mechanism for such an effect is plastic deformation of the hot wire under heavy loads, causing the wire to permanently increase in length. Alternatively, the large forces on the wooden blocks due to the high tension in the wire could cause the blocks (and hence the ends of the wire) to shift closer together. Following the prescription of Ref.~[18], we randomized the order of the trials in each experiment to minimize measurement biases due to these effects. Randomized trials were altered to avoid repetition of the same values of $m$ and $\dT$ on successive trials.

The goal of the first experiment was to empirically determine the minimum load mass $m$ that was still sufficiently heavy that the wire's mass could be neglected. To this end, we measured $H_0$ as a function of $m$ at room temperature. The massless wire model is valid when the quantity $C=m/H_0^3$ is constant with respect to $m$, \emph{i.e.}, when the fluctuations between neighboring data points are smaller than the measurement uncertainty. As can be seen in Fig.~\ref{fig:k}, such is the case when $m\gtrsim 0.7$~kg, implying that the wire's mass was negligible in this regime. The spring constant $k$ was determined from a weighted average of the data in this regime; Eq.~(\ref{eq:k}) gives $k=74(6)$~N/cm, where the quantity in parentheses is the uncertainty in the last digit. The measured spring constant is lower than what one might expect based on nichrome's elastic modulus: $EA/L\approx 200$~N/cm, where $E\approx2\times10^{11}$~N/m$^2$ is nichrome's elastic modulus and $A\approx0.2$~mm${}^2$ is the wire's cross-sectional area.~\cite{Giancoli1991} This discrepancy may be due to kinks in the wire which straighten elastically under tension. In this first experiment, during which all the data were collected at room temperature, there was no evidence of plastic deformation of the wire or shifting of the wooden blocks. Furthermore, we did not observe a dependence of $C$ on $m$ for heavy loads, indicating that Hooke's Law is valid for masses up to at least 1.1~kg.

The purpose of the second experiment was to measure nichrome's thermal expansion coefficient $\alpha$. We measured $H$ as a function of $\dT$ and used Eq.~(\ref{eq:alphaEM}) to determine $\alpha$, using a 0.7~kg load to ensure that the wire's mass could be neglected in our analysis. A weighted average of the data in Fig.~\ref{fig:alpha} yields $\alpha=17.1(1.3)$~\units, which is consistent with the accepted value of nichrome's thermal expansion coefficient, 17.3~\units.~\cite{Shackelford2001} The precision with which we were able to measure $\alpha$ was limited by the uncertainty in measurement of the displacement of the load and the room-temperature length of the wire (Table~\ref{tab:budget}). Both of these sources of uncertainty are due to kinks in the wire.

\FIGX

For fixed $\dT$, we observed a systematic increase in $H$ by about 5\% over the course of the experiment, suggesting that the wire was undergoing plastic deformation. A similar pattern was observed in the room temperature displacement $H_0$, which we measured various times throughout the second experiment. We suspect that kinks in the wire were plastically straightened at high temperatures, a process that would increase the effective spring constant of the wire relative to that observed in the first experiment. Indeed, the values of $H_0$ measured in this experiment correspond to $k=140(50)$~N/cm. The impact of measurement bias due to plastic straightening of the kinks did not significantly affect the accuracy of our experiment. The good agreement between the measured and accepted values of $\alpha$ indicate that the elastic, massless wire approximation is still valid for the stiffer wire.

To determine quantitatively whether or not elastic stretching of the wire is negligible in our experiment, we compare our measurement precision to the fractional correction to the stiff wire model due to the wire's elasticity, \emph{i.e.}, the dimensionless parameter $X$ given in Eq.~(\ref{eq:X}). We find that $X$ varies from 0.5 when $\dT=110$~K to 0.1 when $\dT=530$~K (Fig.~\ref{fig:X}). Because $X$ is larger than our measurement precision of about 0.03, elastic stretching is non-negligible in our experiment.

\section{Future Directions}

Using the thermal expansion experiment described here, we explored the regime of applicability of three models of a wire: stiff and massless, elastic and massless, and elastic and massive. We employed empirical and analytical techniques to develop quantitative conditions for when the mass and elasticity of the wire can be neglected and demonstrated that, for our experiment, the wire's elasticity cannot be neglected. We achieved a measurement of nichrome's coefficient of linear thermal expansion that was consistent with the accepted value at the 8\% level. The dominant sources of uncertainty in our experiment were measurement of the change in height of the load and the room-temperature length of the wire, likely due to kinks in the wire.

One potential method for improving measurement of the load's displacement involves turning the system into a pendulum. By tapping the load and causing it to oscillate such that its displacement is the lever arm of the pendulum, $H$ can be inferred from the period of oscillation. Of course, this technique would require probing the limits of new models, \emph{e.g.}, simple and physical pendulums.

Finally, we note that this experiment is an attractive candidate for teaching and learning about the interplay of measurement uncertainty and models. The simple design of the apparatus and the introductory nature of the corresponding math and physics make these concepts accessible even to beginning students. The apparatus is currently being used in this capacity in a course on measurement designed and taught by the Berkeley Compass Project. Future work will focus on the effectiveness of this experiment as a teaching tool.

\acknowledgments The authors acknowledge helpful discussions with Joel Corbo, Brian Estey, Nathan Leefer, Jenna Pinkham, and William Semel. This work was supported by the Berkeley Compass Project and the Associated Students of the University of California through the Educational Enhancement Fund. DRDF and GZI were supported by the National Science Foundation under grant PHY-1068875.

\end{document}